\magnification=\magstep2
\raggedbottom
\input psfig.sty

\font\lrom=cmr8
\font\srom=cmr9

\def\lsim{\lower.5ex\hbox{$\; \buildrel < \over \sim \;$}} 
\def\gsim{\lower.5ex\hbox{$\; \buildrel > \over \sim \;$}}  
\def\reference{\noindent \hangafter=1 \hangindent=1.0\parindent}

\baselineskip=0.8 truecm

\hsize=17 truecm
\vsize=25 truecm
\hoffset=-0.4truecm
\voffset=-1.2truecm

\nopagenumbers

\null
\vskip 3.5 truecm

\centerline{LOW ENERGY COSMIC RAY INTERACTIONS IN ASTROPHYSICS}
\vskip 2truecm
\centerline{Reuven Ramaty}
\centerline{Laboratory for High Energy Astrophysics}
\centerline{Goddard Space Flight Center, Greenbelt, MD 20771, 
USA}

\vskip 1.5 truecm

\centerline{Abstract}

Gamma ray emission lines resulting from accelerated particle 
bombardment of ambient gas can serve as an important 
spectroscopic tool for abundance determinations. The method is 
illustrated by considering the gamma ray line emission observed 
from solar flares. The observation of similar gamma ray lines 
from Orion suggests the existence of large fluxes of 
low energy Galactic cosmic rays. The role of these cosmic rays 
in the nucleosynthesis of the light isotopes is discussed. 

\vskip 1 truecm

\centerline{To be published in}
\vskip 0.1truecm
\centerline{Proceedings of the ``Sun and Beyond" Conference, 1996}
\centerline{Held in Ho Chi Minh City, Vietnam, October 1995}

\eject

\vskip 0.3 truecm
\centerline{\bf Introduction}
\vskip 0.2 truecm

The interactions of accelerated particles with ambient matter 
produce a variety of gamma ray lines following deexcitations in 
both the nuclei of the ambient medium and the accelerated 
particles. Astrophysical deexcitation line emission produced by 
accelerated particle interactions has so far been observed from 
solar flares$^{1,2)}$ and the Orion molecular cloud 
complex$^{3)}$. Recent general reviews of astrophysical gamma 
ray line emission are available$^{4,5)}$. The solar gamma ray 
line observations have many applications, including the 
determination of solar atmospheric abundances$^{6,7,8)}$. The 
Orion observations, even though much less detailed, have 
nonetheless revealed the existence of large fluxes of low 
energy cosmic rays in this nearest region of recent star 
formation [e.g. ref.$^{9)}$]. If such cosmic rays are also 
present at other sites in the Galaxy, then low energy Galactic 
cosmic rays may play an important role in the nucleosynthesis 
of the light isotopes $^6$Li, $^9$Be, $^{10}$B and $^{11}$B 
[refs.$^{10,11,12)}$].

In the present paper we review these topics, referring the 
reader for more details to the papers referenced above.

\vskip 0.3 truecm
\centerline{\bf Solar Gamma Ray Spectroscopy}
\vskip 0.2 truecm

The solar flare gamma ray data is now sufficiently detailed to 
allow the conduct of a meaningful gamma ray spectroscopic 
analysis of the ambient solar atmosphere. The key is provided 
by the narrow line emission produced by accelerated protons and 
$\alpha$ particles interacting with ambient C and heavier 
nuclei. Owing to their narrower widths, these lines can be 
distinguished from the broader lines produced by accelerated C 
and heavier nuclei interacting with ambient H and He. The 
intensities of the narrow lines depend on the heavy element 
abundances and thus can be used to determine these abundances. 
Strong narrow line emission at 4.44, 6.13, 1.63, 1.37, 1.78, 
and 0.85 MeV, resulting from deexcitations in $^{12}$C, 
$^{16}$O, $^{20}$Ne, $^{24}$Mg, $^{28}$Si and $^{56}$Fe, 
respectively, has been observed from many flares. The most 
recent results, observed with the Solar Maximum Mission (SMM) 
from 19 flares$^{2)}$, allow the determination$^{7,8)}$ of the 
abundance ratios C/O, Mg/O and Mg/Ne for all 19 flares, Si/O 
for 14 flares and Fe/O for 12 flares.

Unlike atomic spectroscopy, nuclear spectroscopy does not 
require the temperature and ionic state of the ambient gas, 
neither of which are always well known. On the other hand, 
abundance determinations by nuclear spectroscopy do require 
information on the spectrum of the accelerated particles. For 
the SMM flare analysis$^{7,8)}$ the accelerated particle 
spectra were constrained by using the 1.63 MeV 
$^{20}$Ne-to-6.13 MeV $^{16}$O and the 2.22 MeV neutron 
capture-to-4.44 MeV $^{12}$C line fluence ratios, both of which 
are strong functions of the particle spectrum. 

As in other solar atmospheric abundance studies [e.g. 
ref.$^{13)}$], it is useful to distinguish two groups of 
elements depending on their first ionization potential (FIP): 
low FIP ($<$10 eV) elements (Mg, Si and Fe) and high FIP ($>$11 
eV) elements (C, O and Ne). The enhancement of low FIP-to-high 
FIP element abundance ratios in the corona relative to the 
photosphere is well established from atomic spectroscopy and 
solar energetic particle observations [e.g. ref.$^{13)}$]. 
Analysis of the gamma ray data has led to the following 
conclusions$^{7,8)}$:

(i) For the high FIP elements C and O, the derived abundance 
ratio (by number) is 0.35$\lsim$C/O$\lsim$0.44 [Fig.~1, from 
ref.$^{8)}$]. This range is more consistent with C/O = 
0.43$\pm$0.05 [ref.$^{14)}$] than with C/O=0.48$\pm$0.1  
[ref.$^{15)}$]. But taking into account the large uncertainty 
of the latter, there is no real discrepancy. Furthermore, a 
single value of C/O is consistent with the data for all 19 
flares, implying that C/O could have the same value throughout 
the gamma ray production region. This is in fact not surprising 
given that C/O is essentially the same in the photosphere and 
corona.

(ii) For another pair of high FIP elements, O and Ne, the gamma 
ray data is in better agreement with Ne/O=0.25 than with the 
commonly adopted photospheric and coronal value of 0.15. Such a 
low Ne/O could only be accommodated by a very steep accelerated 
particle spectrum which would take advantage of the very low 
threshold for the excitation of the 1.63 MeV level of 
$^{20}$Ne. The implied particle spectra, however, are too steep 
to produce sufficient neutrons to account for observations of 
the 2.22 MeV neutron capture line. In addition, the energy 
contained in ions with such steep spectra would be inconsistent 
with the overall flare energetics. Some EUV and X-ray 
observations$^{16,17,18)}$ also support a Ne/O higher than 0.15.

\vskip -2.0 truecm
\psfig{figure=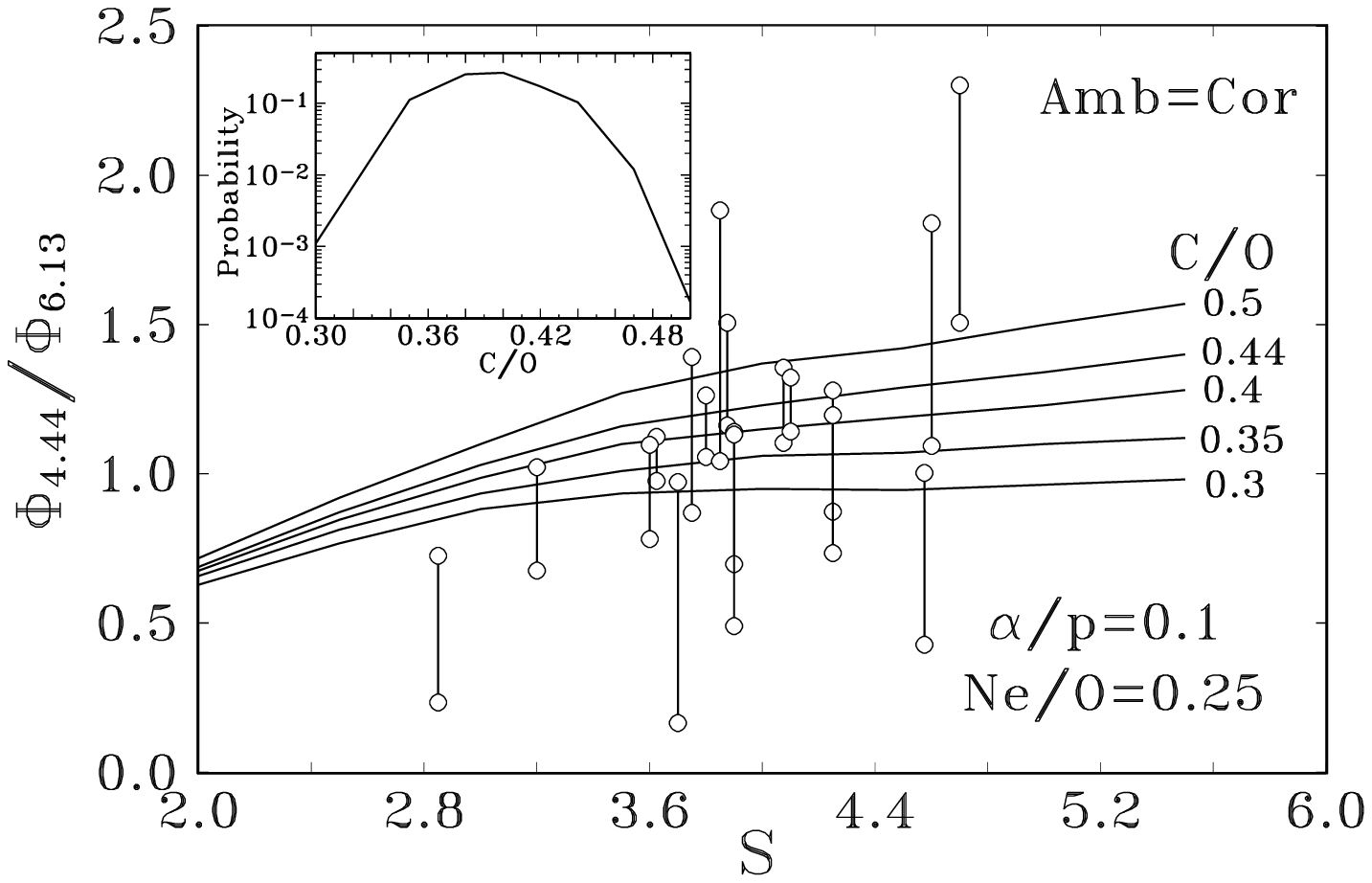,height=18truecm,width=24truecm}
\vskip -7.0truecm

$$\vbox{

\baselineskip=0.5truecm \lrom 

\noindent Fig.~1. Observed$^{2)}$ and calculated$^{8)}$ 
$^{12}$C and $^{16}$O deexcitation line ratios [from 
ref.$^{8)}$]. $S$ is the power law spectral index of the 
accelerated particles; $\alpha/p$ is the $\alpha$ particle to 
proton ratio for these particles; C/O is the C to O abundance 
ratio in the ambient medium which was allowed to vary; the 
other ambient abundance ratios were kept fixed at coronal 
values. Ne/O=0.25 is the ambient Ne/O abundance ratio used to 
derive the value of $S$ for each flare [see (ii) above]. The 
insert gives the $\chi^2$ derived probabilities corresponding 
to the fits of the various calculated curves to the data; the 
best fit is achieved at C/O=0.4.

}$$

(iii) To avoid values of Ne/O larger than 0.3 the accelerated 
particle energy spectra should be at least as steep as an 
unbroken power law down to about 1 MeV/nucl. For such power 
laws, the energy contained in the ions for the 19 analyzed 
flares ranges from about 10$^{30}$ to well over 10$^{32}$ ergs, 
and is thus comparable or even exceeds the energy contained in 
the nonrelativistic electrons that produce the hard X-rays in 
solar flares. Prior to these recent gamma ray analyses, it was 
widely believed that a large fraction of the released flare 
energy is contained in nonrelativistic electrons.  

(iv) Considering the abundance ratios between elements of 
different FIP groups [Fig.~2, from ref.$^{8)}$], both Mg/O and 
Mg/Ne show evidence for variability from flare to flare at 
about the 3$\sigma$ level. For Mg/O this variation is confined 
to a range around the coronal value of 0.2 and does not go down 
to the photospheric value of 0.045. For Si/O and Fe/O the 
variations are also confined to a range around their respective 
coronal values. The fact that the low FIP-to-high FIP abundance 
ratios derived from gamma ray spectroscopy are enhanced 
relative to their respective photospheric values shows that the 
gamma ray production region lies above the photosphere.

\vskip 0 truecm
\psfig{figure=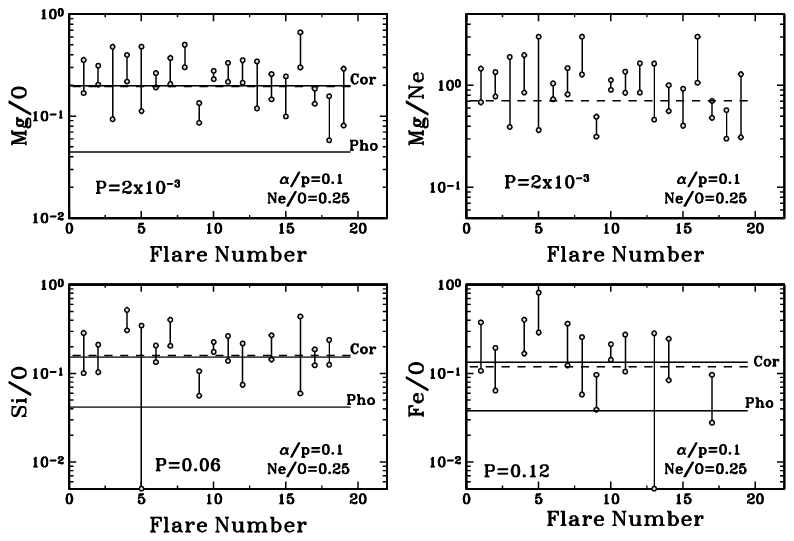,width=6.5in,bbllx=0bp,bblly=0bp,bburx=270bp,bbury=190bp,clip=t}
\vskip -6 truecm

$$\vbox{

\baselineskip=0.5truecm \lrom \noindent Fig.~2. Derived Mg/O, 
Mg/Ne, Si/O and Fe/O abundance ratios. The solid lines are 
coronal and photospheric abundances; the dashed lines are the 
best fitting abundances. For Mg/O and Mg/Ne there are data for 
all 19 flares; however, for Si/O and Fe/O the data are more 
limited. 

}$$

\centerline{\bf Orion and the Origin of the Light Isotopes}

The discovery$^{3)}$ of gamma ray line emission from Orion with 
the COMPTEL instrument on the Compton Gamma Ray Observatory 
(CGRO) and the implied existence of large fluxes of low energy 
Galactic cosmic rays, has led to renewed discussions on the 
origin of the light elements. Some of the important 
implications of the Orion gamma ray observations are the 
following [see ref.$^{9)}$ for a more detailed summary):

\vskip -2truecm
\psfig{figure=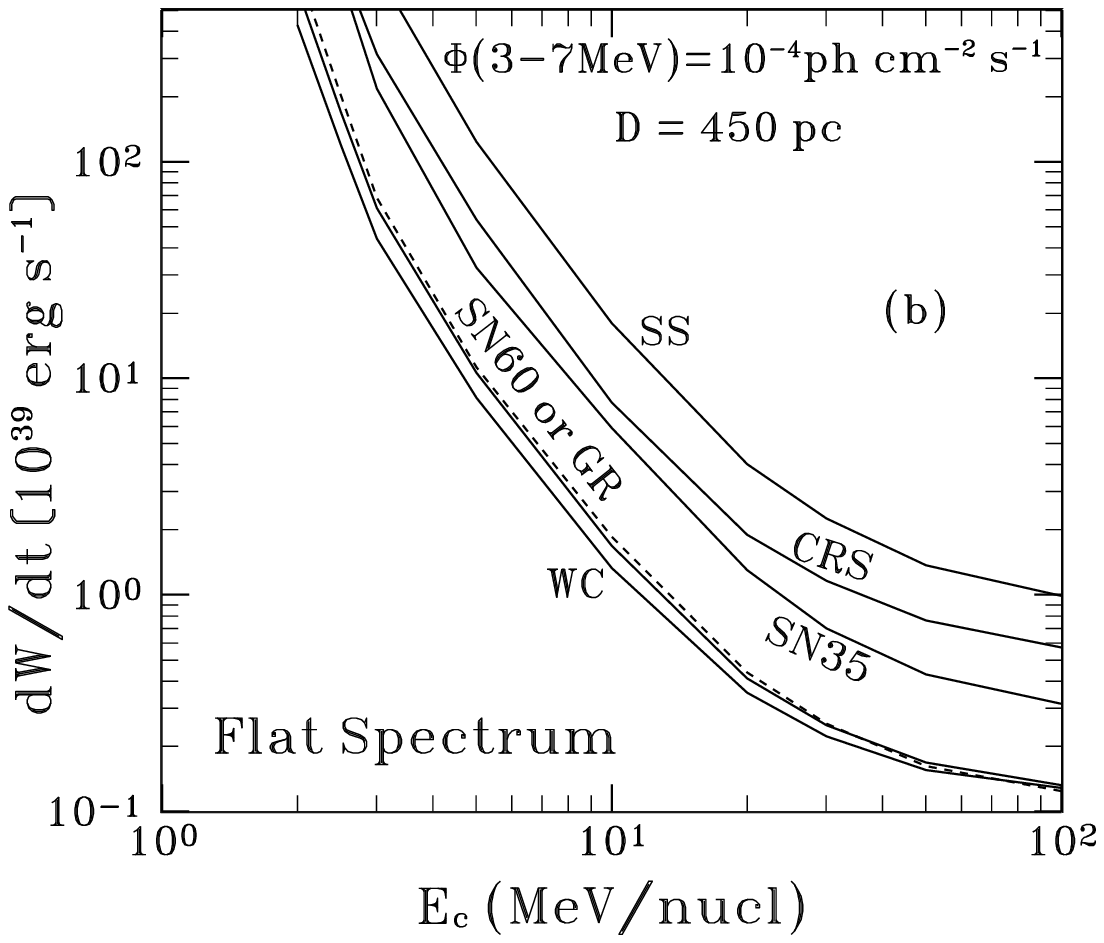,width=6.5in}
\vskip -6.5truecm

$$\vbox{\baselineskip=0.5truecm \lrom 

\noindent Fig.~3. Deposited power that accompanies the 
production of the observed gamma ray line emission from Orion 
for energy spectra characterized by a typical energy $E_c$ and 
various compositions for the accelerated particles. The SN60 
(ejecta of a supernova from massive star progenitors), GR 
(acceleration of the products of the breakup of interstellar 
dust) and WC (winds of Wolf-Rayet stars of spectral type WC) 
compositions have much lower proton and $\alpha$ particle 
abundances relative to heavier nuclei than do the SS (solar 
system) and CRS (cosmic ray source) compositions. The deposited 
power is normalized to the observed 3--7 MeV flux seen from 
Orion. For the SN, GR and WC compositions, and $E_c$=30 
MeV/nucl, the deposited power is about $2.5\times10^{38}$ ergs 
s$^{-1}$, which integrates in 10$^5$ years to about 
$7.5\times10^{50}$ ergs in low energy cosmic rays. These cosmic 
rays could have been accelerated by an 80,000 year old 
supernova in the OB association whose direction coincides with 
the centroid of the Orion gamma ray line emission$^{12)}$; this 
supernova is thought$^{20)}$ to have produced the 
Orion-Eridanus bubble seen in soft X-rays. For much lower 
values of $E_c$ and the other compositions, which are not 
depleted in protons and $\alpha$ particles, the power required 
to produce the observed gamma ray line emission is much higher.

}$$

(i) The typical energies of the low energy cosmic rays are 
around tens of MeV/nucleon. For much higher energies gamma ray 
production via pion decay would lead to gamma ray fluxes which 
would exceed the observed$^{19}$ flux with the EGRET 
instrument on CGRO. Much lower energies would render the gamma 
ray line production energetically very inefficient [Fig.~3, 
from ref.$^{12)}$].

(ii) The observed gamma ray line emission is more likely to be 
produced by accelerated particles which are strongly depleted 
in protons and $\alpha$ particles relative to heavier nuclei 
than by particles with a more 'conventional' composition. While 
in principle this conclusion could follow from the widths of 
the observed gamma ray lines, in practice the COMPTEL data are 
still not of high enough quality to allow this 
distinction$^{21)}$. The conclusion that the proton and 
$\alpha$ particle abundances are suppressed has come from 
arguments of energetics: the absence of protons and $\alpha$ 
particles minimizes the deposited power that accompanies the 
production of the gamma ray line emission [refs.$^{11,12)}$, 
Fig.~3]. This suppression could be the consequence of the 
particle injection process prior to the acceleration itself. 
The proposed injection sources are the winds of Wolf Rayet 
stars$^{21)}$, the ejecta of supernovae from massive star 
progenitors$^{10,12)}$, and the pick up ions resulting from the 
breakup of interstellar grains$^{11,12)}$.

It has been known for over two decades that the relativistic 
Galactic cosmic rays (GCR) may have produced$^{22,23)}$ the 
observed solar system abundances of $^6$Li, $^9$Be and 
$^{10}$B. These cosmic rays, however, cannot account for the 
abundances of $^7$Li and $^{11}$B. It is believed$^{24)}$ that 
most of the Galactic $^7$Li is produced in stars and in the Big 
Bang. Recent measurements$^{25)}$ of the boron isotopic ratio 
in meteorites yielded $^{11}$B/$^{10}$B values in the range 
3.84 -- 4.25 which exceed the calculated GCR value by a factor 
of about 1.5. The implications of the Orion gamma ray 
observations on the origin of the light isotopes have been 
considered$^{10,11,12)}$. The advantages of producing the light 
isotopes with 'Orion-like' low energy cosmic rays are the 
following: 

(i) Low energy cosmic rays can produce B such that 
$^{11}$B/$^{10}$B$\gsim$4. We illustrate this in Fig.~4 
[original to the current paper but based on calculations 
similar to those presented previously$^{12)}$]. As we have 
seen, arguments of energetics for Orion favor low energy cosmic 
rays with $E_c$ around 30 MeV/nucl. At such energies the excess 
$^{11}$B results mostly from $^{12}$C via the reactions 
$^{12}$C(p,pn)$^{11}$C and $^{12}$C(p,2p)$^{11}$B which have 
lower thresholds than the reaction $^{12}$C(p,2pn)$^{10}$B. At 
higher energies, and in reactions with $^{16}$O, the B isotopic 
ratio is significantly lower. 

(ii) As for the gamma ray production in Orion, if low 
energy cosmic rays play a role in light element production, 
then they are probably depleted in protons and $\alpha$ 
particles. The $\alpha$ particle depletion is necessary in 
order not to overproduce $^6$Li; the proton depletion allows a 
linear dependence of the Be and B abundances on the Fe 
abundance in stars of various ages. If the low energy cosmic 
rays are poor in protons and $\alpha$ particles they will 
produce Be and B only from the breakup of accelerated C and O 
in interactions with ambient H and He; in this case both the 
target and projectile abundances could remain constant, leading 
to a linear growth of the Be and B abundances. On the other 
hand, the GCR would produce much of the isotopes from the 
breakup of C and O in the ambient medium whose abundances 
increase with time, leading to a quadratic growth.

\psfig{figure=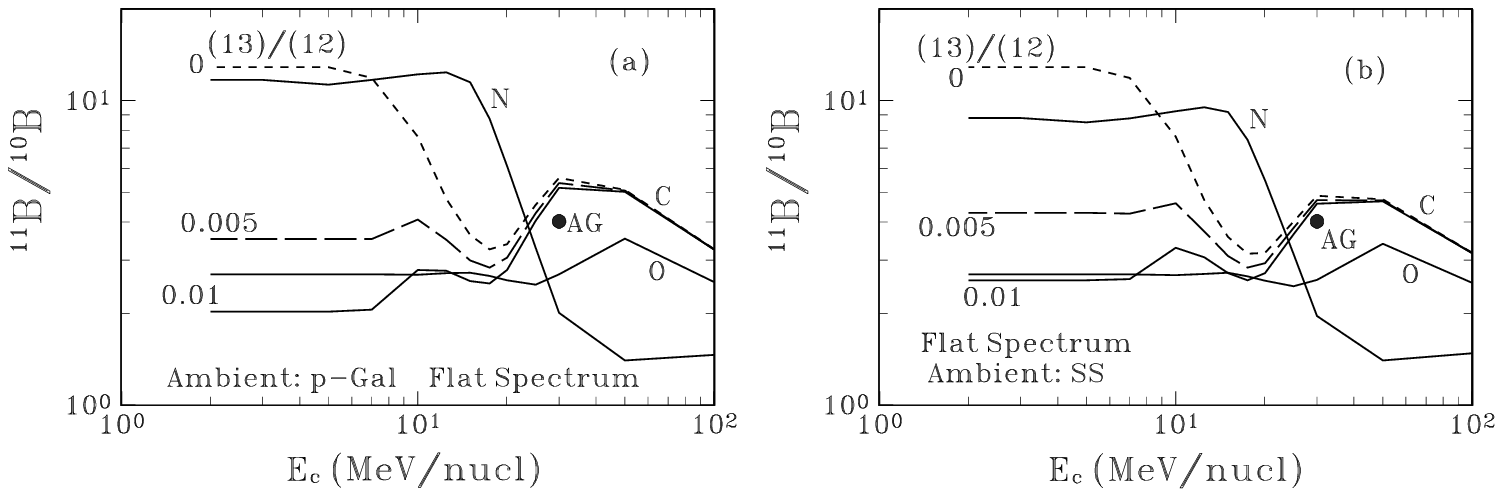,width=6.5in,bbllx=0bp,bblly=0bp,bburx=467bp,bbury=190bp}
\vskip -4.5 truecm

$$\vbox{\baselineskip=0.5truecm \lrom 

\noindent Fig.~4. B isotopic ratio produced separately by 
accelerated C, N or O. In panel (a) (proto-Galactic) the 
ambient medium has no metals and He/H=0.075 by number; the 
solar system composition in panel (b) has He/H=0.1 and 
photospheric metal abundances. In both panels there are 
calculations for 3 values for the isotopic ratio of the 
accelerated C. The AG value is the meteoritic B isotopic 
ratio$^{14)}$. Above about 20 MeV/nucleon the meteoritic ratio 
requires the presence of C in the accelerated particles.

}$$

(iii) The arguments of energetics are also relevant for the 
light element production. Fig.~5 [from ref.$^{12)}$] shows the 
total energy in accelerated particles that is required to 
produce 100 M$_\odot$ of B, which is the estimated Galactic 
inventory of this light element. We see that $W$  increases 
rapidly as $E_c$ decreases and that $W$ is minimized by the 
compositions depleted in protons and $\alpha$ particles. 
Considering the SN60 case, supernovae from massive star 
progenitors are excellent candidates for close interaction with 
molecular clouds since they evolve quite rapidly and thus 
explode before they have moved very far from the cloud where 
they were formed. The maximum available energy in low energy 
cosmic rays from all $>$60 M$_{\odot}$ Galactic supernovae is 
roughly 2$\times$10$^{58}$ ergs, shown as the horizontal bar in 
Fig.~5. This estimated maximum assumes that all of the Galactic 
$^{56}$Fe of 6$\times$10$^7$ M$_{\odot}$ is produced by Type II 
supernovae from $>$8 M$_{\odot}$ progenitors, each producing an 
average of 0.1 M$_{\odot}$ of $^{56}$Fe. This gives a total of 
6$\times$10$^8$ Type II supernovae in the age of the Galaxy. 
Scaling of the supernova rate by an initial mass function 
proportional to $M^{-2.7}$, then gives a maximum of 
2$\times$10$^7$ Type II supernovae from $>$60$M_\odot$ 
progenitors, each of which could have as much as $\sim$ 
10$^{51}$ erg of mechanical energy available for low energy cosmic 
ray acceleration. This energetic argument thus gives further 
support to the previous conclusions that the light elements are 
produced by low energy cosmic rays of typical energies around 
30 MeV/nucl and depleted in protons and $\alpha$ particles.

\psfig{figure=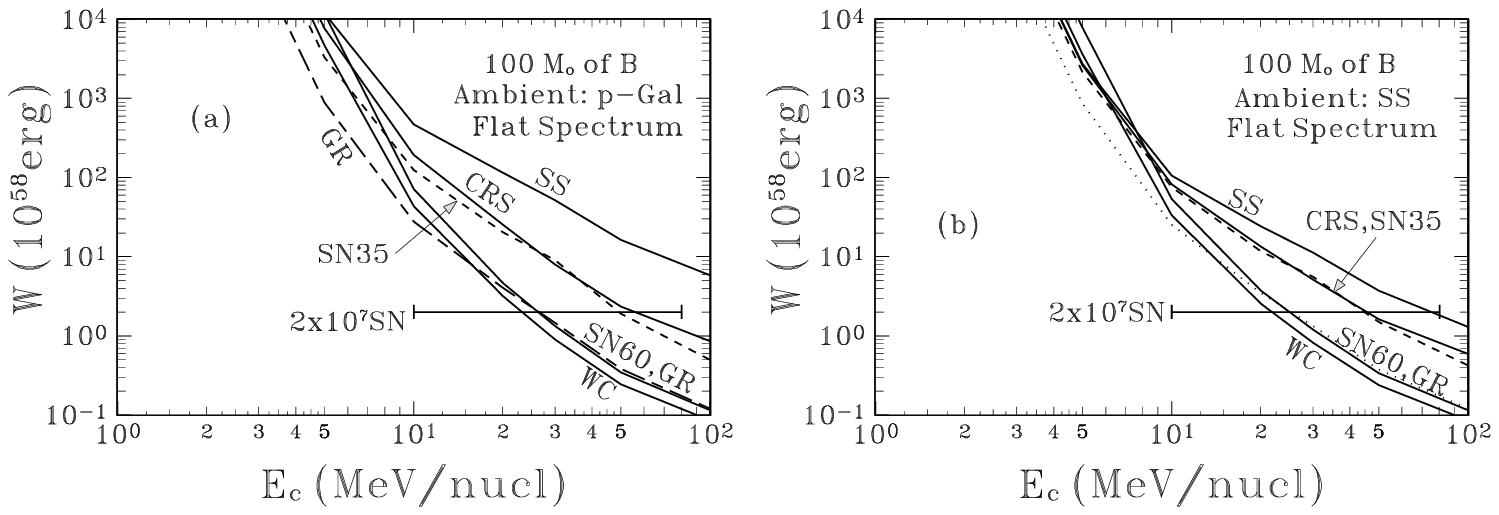,width=6.5in,bbllx=0bp,bblly=0bp,bburx=467bp,bbury=190bp}
\vskip -4.5 truecm

$$\vbox{\baselineskip=0.5truecm \lrom 

\noindent Fig.~5. The energy deposited during the production of 
100 M$_\odot$ of B. The assumed compositions for the 
accelerated particles are the same as those in Fig.~3. As in 
Fig.~4, panels (a) and (b) are for the proto-Galactic and solar 
system ambient medium compositions, respectively. 

}$$

\vskip 0.3 truecm
\centerline{\bf Summary}
\vskip 0.2 truecm

We have used solar flare observations to demonstrate the new 
technique of abundance determination based on gamma ray line 
emission produced by accelerated particle bombardment. We have 
also reviewed some of the implications of similar gamma ray 
line emission observed from Orion, including the production of 
light elements by low energy cosmic rays. 

For such an origin for the light elements, limitations on the 
total available energy in low energy cosmic rays imply that the 
production takes place at relatively high particle energies, 
around 30 MeV/nucl. Such low energy cosmic rays can reproduce 
the meteoritic B isotopic ratio of 4, but they require an 
accelerated C/O on the order of the observed solar system ratio 
($\sim$0.5). If in the early Galaxy C/O in the ejecta of Type 
II supernovae from massive star progenitors (which are thought 
to be the sources of the postulated cosmic rays) is much lower, 
then the B isotopic ratio in low metallicity stars should be 
significantly lower than the measured meteoritic value. 

\vskip -0.3truecm 

$$\vbox{\baselineskip=0.55truecm \srom

\vskip 0.3 truecm
\centerline{\bf References}
\vskip 0.2 truecm

\reference 1. Chupp, E. L. 1990, Physica Scripta, T18, 15

\reference 2. Share, G. H. \& Murphy, R. J. 1995, ApJ, 452, 933

\reference 3. Bloemen, H. et al. 1994, A\&A, 281, L5

\reference 4. Ramaty, R. \& Lingenfelter, R. E. 1995, in The 
Analysis of Emission Lines, eds. R. E. Williams and M. Livio, 
(Cambridge: Cambridge Univ. Press), 180

\reference 5. Prantzos, N. 1996, A\&A, in press

\reference 6. Murphy, R. J., Ramaty, R., Kozlovsky, B., \& Reames, D. 
V. 1991, ApJ, 371, 793

\reference 7. Ramaty, R., Mandzhavidze, N., Kozlovsky, B., \& 
Murphy, R. J. 1995, ApJ, 455, L193

\reference 8. Ramaty, R., Mandzhavidze, N., \& Kozlovsky, B., 
1996, in High Energy Solar Physics, R. Ramaty, N. Mandzhavidze, 
X.-M. Hua, eds. (AIP: New-York), in press

\reference 9. Ramaty, R. 1996, A\&A, in press

\reference 10. Cass\'e, M., Lehoucq, R., \& Vangioni-Flam, E., 1995, 
Nature, 373, 318

\reference 11. Ramaty, R., Kozlovsky, B., \& Lingenfelter, R. 
E. 1995, Annals. N. Y. Acad. of Sci. (17th Texas Symposium on 
Relativistic Astrophysics and Cosmology, eds. H. Bohringer, G. 
E. Morfill and J. Trumper), 759, 392

\reference 12. Ramaty, R., Kozlovsky, B., \& Lingenfelter, R. E. 
1996, ApJ, in press (Jan 10)

\reference 13. Meyer, J-P. 1992, in Origin and Evolution of the Elements,
eds. N. Prantzos et al. (Cambridge: Cambridge Univ. Press), 26 

\reference 14. Anders, E. \& Grevesse, N. 1989, Geochim. et Cosmochim. 
Acta, 53, 197

\reference 15. Grevesse, N. \& Noels, A. 1992, in Origin and Evolution
of the Elements, eds. N. Prantzos et al. (Cambridge: Cambridge Univ.
Press), 14

\reference 16. Saba, J. L. R. \& Strong, K. T. 1993, Adv. Sp. Res., 13 (9)391

\reference 17. Schmelz, J. T. 1993, ApJ, 408, 373

\reference 18. Widing, K. G. \& Feldman U. 1995, ApJ, 442, 446

\reference 19. Digel, S. W., Hunter, S. D., \& Mukherjee, R. 1995, 
ApJ, 441, 270

\reference 20. Burrows, D. N., Singh, K. P., Nousek, J. A., Garmire, G. 
P., \& Good, J. 1993, ApJ, 406, 97

\reference 21. Ramaty, R., Kozlovsky, B., \& Lingenfelter, R. E. 
1995, ApJ, 438, L21

\reference 22. Meneguzzi, M., Audouze, J., and Reeves, H. 1971, 
A\&A, 15, 337

\reference 23. Mitler, H. E. 1972, Ap\&SS, 17, 186

\reference 24. Reeves, H. 1994, Revs. Modern Physics, 66, 193

\reference 25. Chaussidon, M. \& Robert, F. 1995, Nature, 374, 337

}$$

\bye